\documentclass[
    ,final            
  ]
  {aipproc}

\layoutstyle{6x9}

\begin{document}

\title{Baryons and Their Halos}


\author{Stacy McGaugh}{
  address={Department of Astronomy, University of Maryland, College Park, MD 20742-2421 USA}
}

\begin{abstract}
Galaxies are composed of baryonic stars and gas embedded in dark matter halos. 
Here I briefly review two aspects of the connection between baryons and their halos.
(1) The observed baryon content of galaxies falls short of the cosmic baryon fraction
by an amount that varies systematically with mass.  Where these missing baryons
now reside is unclear.  (2) The characteristic acceleration
in disk galaxies correlates strongly with their baryonic mass surface density.  This implies
a close coupling between the gravitational dynamics, which is presumably dominated by
dark matter, and the purely baryonic components of galaxies.
\end{abstract}

\maketitle

\section{Introduction}

Our current cosmological paradigm envisions cosmic structures like galaxies being embedded 
in extended halos of cold dark matter.  Enormous progress has been made in simulating the
growth of large scale structure in the dark component \cite{milleneum}.  Forming actual galaxies
in simulations, by which I mean the baryonic components we can observe, has proven to be rather
more difficult.

A wide gulf exists between our theoretical notions about how galaxies should be and our observational
picture of how they are.  Most of the problems stem from the connection between the dark matter and
the harder to simulate baryons.  Here I discuss two of the outstanding issues.

\section{The Baryon Content of Dark Matter Halos}

The early universe was a highly uniform mix of photons, dark, and baryonic matter
with a cosmic baryon fraction $f_b = 0.17$ \cite{WMAP5}.  As dark matter halos form, 
baryons are supposed to condense within them to form luminous galaxies \cite{whiterees}.
I would naively expect the cosmic fraction of baryons to be retained by each and every dark 
matter halo.  We can therefor test this hypothesis by making an accounting of the baryon
content of individual dark matter halos.

\begin{figure}[t]
  \includegraphics[width=\textwidth]{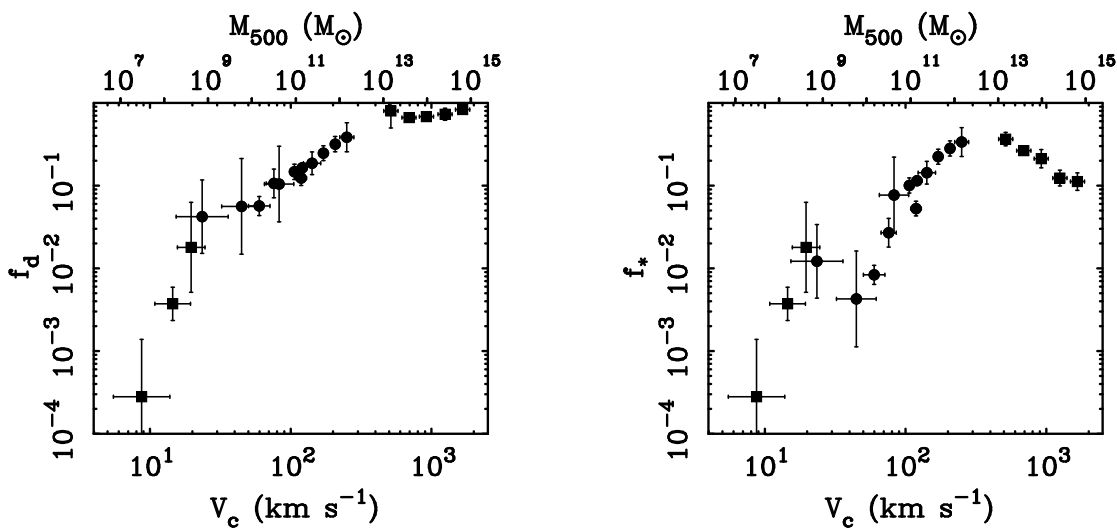}
  \caption{The fraction of the expected baryons that are detected [left: ${f_d = M_b/(f_b M_{500})}$] 
and the fraction converted into stars [right: ${f_* = M_*/(f_b M_{500})}$] as per \cite{M10}.
The detected baryon fraction increases monotonically with mass while the stellar fraction peaks
between $\mathrm{M}_{500} = 10^{12}$ and $10^{13}\;\mathrm{M}_{\odot}$.
\label{fdfstV}
}
\end{figure}

Such an inventory has recently been assembled \cite{M10}.  The basic result, which has
long been apparent \cite{aas}, is that most
of the expected baryons are missing (Fig.~\ref{fdfstV}).  Not only do we suffer form a dynamical
missing mass problem (for which we invoke CDM), but we also suffer from a missing baryon
problem.  This is distinct from the global missing baryon problem in which the sum of known
baryons falls short of the density expected from Big Bang Nucleosynthesis \cite{fukugita}.  
Here, individual objects are missing baryons on a halo by halo basis.

An obvious question, then, is where are all these missing baryons?  The most popular idea seems
to be that they are in undetected hot gas that has failed to cool and condense into galaxies or
was blown away by supernova feedback.  The first place one would expect to
find such baryons is in the halo: away from the central galaxy but still mixed in with the dark matter
halo.  This does not appear to be the case, however, as direct searches do not find the required
mass \cite{bregman,wheregas}.  

It may be possible that feedback has liberated many baryons entirely form their parent halos.
Indeed, the retained baryon fraction strongly correlates with potential well depth (Fig.~\ref{fdfstV}).
It seems natural therefor to invoke mass-dependent feedback in order to explain the observed
trend.  However, getting the numbers right remains a huge challenge.  Even the most sophisticated simulations involving feedback \cite{governato,cutesim}, beautiful as they are, fail by an order of magnitude.
In the particular case of \cite{governato,cutesim}, roughly half the baryons are retained at a scale where only
about 5\% are observed.

\section{Halo-Baryon Coupling}

Dark matter provides the dominant total mass of a galaxy.  Within the optical radius, dark matter
appears to dominate in some types of systems, like low surface brightness disks \cite{dBMH}
and Milky Way satellites \cite{walker}.  In high surface brightness spiral and elliptical galaxies,
baryons appear to play a dynamically important role in the inner parts \cite{edo,romanowsky}.
However, there is a deeper issue than the degree to which baryons contribute to the 
gravitational potential \cite{dmass}.  It is
understanding the palpable coupling between the two components \cite{sancisi,M04}. 

\begin{figure}[t]
  \includegraphics[width=\textwidth]{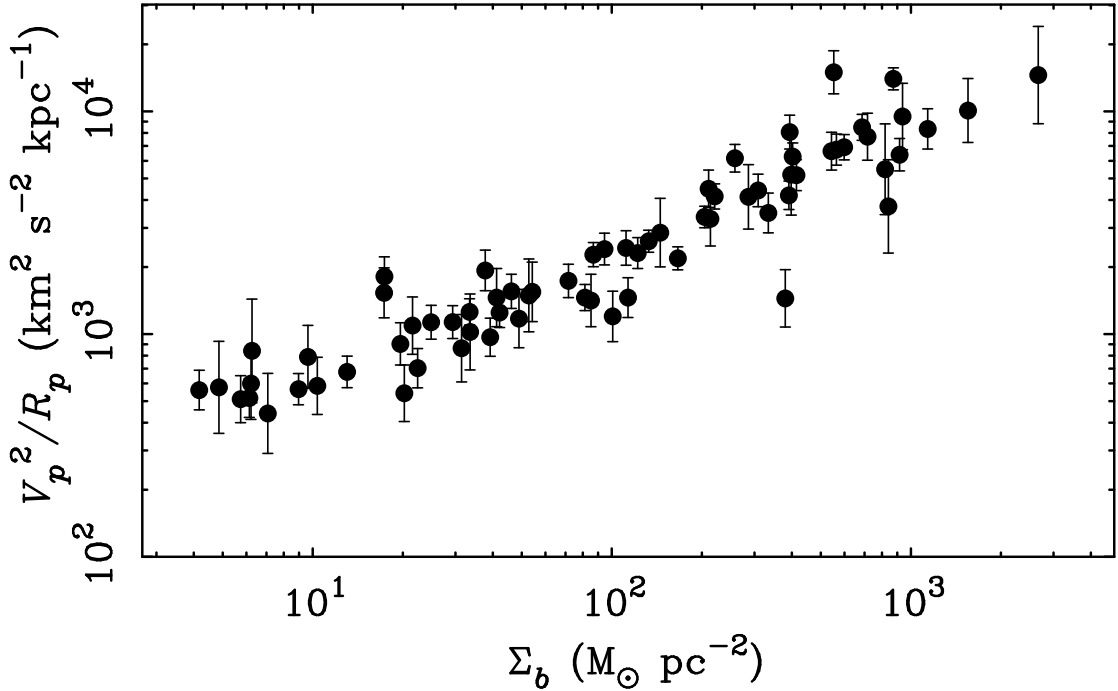}
  \caption{The characteristic acceleration of disk galaxies as
  a function of their baryonic surface mass density. The strong correlation
  is a clear indication of a coupling between baryonic
  and dark components:  the two ``know'' about each other \cite{sancisi,M04}. 
    \label{XiSd}}
\end{figure}

Fig.~\ref{XiSd} plots the characteristic acceleration against the baryonic surface density
for a sample of spiral galaxies \cite{M05}.  The characteristic acceleration is taken to be
the centripetal acceleration measured at the point were the contribution of the baryons
peaks.  For a purely exponential disk, this point is at $R_p = 2.2$ scale lengths \cite{CR}.  
No approximation is made here; the peak radius $R_p$ is taken at the actual maximum
of the mass model, including both the bulge and stellar and gaseous disks \cite{prl}.
If dark matter dominates throughout, then the acceleration measured at this point is 
dominated by the halo and should not depend on the baryonic surface density.
   
The characteristic  baryonic surface mass density is the central surface density of the
exponential disk that is equivalent to that which the actual baryonic mass would have if
bulge, stellar, and gaseous disk were all arranged into a single exponential distribution \cite{prl}.
In order to include the gas mass, it is necessary to assign some mass-to-light ratio to the stars.
For specificity, we use the stellar population models of \cite{Bell03} here.
The choice of stellar mass estimator is not critical.  A different choice would merely translate the
abscissa, modulo the gas surface density.  The basic relation is already apparent with 
raw surface brightness \cite{MdB98b}.
  
Contrary to the expectation for complete dark matter domination, the observed \textit{dynamical}
acceleration correlates strongly with observed \textit{baryonic} surface density.  Clearly the baryons do
matter to the dynamics.  Indeed, the baryonic surface density is predictive of the total gravitational
potential due to both dark and baryonic mass.  This is true not only at the specific location illustrated
in Fig.~\ref{XiSd}, but at all radii in disk galaxies \cite{M04}.  Thus Renzo's rule:
\textit{``For any feature in the luminosity profile there is a corresponding feature in the rotation curve
 and vice versa''}  \cite{sancisi}.
 
We remain very far from being able to reproduce the observed coupling between dark and
luminous matter in galaxy formation simulations.  Indeed, this strikes me as a serious
fine-tuning problem for any $\Lambda$CDM model.  In this context,
I find it disturbing that the only theory to have 
correctly predicted the observed behavior \textit{a priori} is MOND \cite{milgrom83}.

\begin{theacknowledgments}
  The work of SSM is supported in part by NSF grant AST-0908370.
\end{theacknowledgments}

\bibliographystyle{aipproc}   % if natbib is available

\end{document}